\documentclass[letterpaper, 10 pt,  conference]{ieeeconf}  

\IEEEoverridecommandlockouts                              
\usepackage{graphics} 
\usepackage{epsfig} 
\usepackage{mathptmx} 
\usepackage{wrapfig}
\usepackage{times} 
\usepackage{amsmath} 
\usepackage{amssymb}  
\usepackage{array,multirow}
\usepackage{adjustbox}
\usepackage{fp}
\usepackage{subcaption}
\usepackage{xcolor}
\usepackage{dblfloatfix}

\title{\LARGE \bf
Dynamic Tracking Biosensors: Unconstrained\\ Detection and Performance Limits
}

\author{Deepak Gopalan and Pradeep R. Nair\\
Department of Electrical Engineering\\Indian Institute of Technology Bombay, Mumbai, India
}

\begin{document}
 \captionsetup[figure]{labelfont={bf},name={Fig.},labelsep=period}

\maketitle
\thispagestyle{plain}
\pagestyle{plain}

\begin{abstract}
Accurate detection of target molecules at low concentration in the presence of high concentration of undesired molecules is a major challenge for End Point ($EP$) assays. Non-specific binding of undesired molecules to receptors limits the minimum detectable concentration of the target significantly. Dynamic tracking ($DT$) of binding and unbinding events allows us to overcome this challenge and provides a remarkable improvement in the minimum detectable target concentration, as demonstrated recently. In this manuscript, we propose a novel unconstrained detection scheme which does not rely on a priori knowledge of the reaction constants. This scheme allows facile back extraction of various critical sensor parameters as well. Further, through a combination of theoretical analysis and detailed statistical simulations, we show that  $DT$ sensors could be several orders of magnitude better than $EP$ biosensors. This work identifies and establishes the functional dependence of critical parameters on the performance of $DT$ sensors and hence could be of broad interest to the community towards further optimization.
\end{abstract}

\section{\textbf{Introduction}}
Most ultra-sensitive assays aim to detect low concentration targets by using receptors which bind specifically to target molecules. Although chosen to be specific, other molecules present in the solution might also bind to the receptors. If the target molecules exist in very low concentrations while the undesired molecules exist in abundance, this problem is further exacerbated. Despite the poor affinity to receptors, the sheer amount of undesired molecules might result in a significant number of receptors binding to these. This not only prevents the target molecules from binding to the receptors, but also causes a faulty inference about the target concentration. Hence, accurately detecting target molecules in the presence of highly abundant undesired molecules is a fundamental challenge. \vspace{-0.7em}\\

Endpoint ($EP$) assays measure the amount of bound receptors after a certain incubation time to detect the target concentration \cite{Squires2008}. Such schemes are often incapable of resolving the issue highlighted above. Dynamic tracking ($DT$) or kinetic assays was proposed recently to overcome this drawback \cite{Sevenler4129}. Here, individual binding-unbinding events happening on the sensor surface is tracked over time and receptor-target binding is distinguished from undesired or non-specific binding (see Fig. 1). Recently developed digital microarrays are capable of performing such single molecule readout with good fidelity. Previous work demonstrated the measurement of DNA molecules up to a concentration of 19fM by applying dynamic tracking to a video feed of a digital microarray \cite{Sevenler4129}. However, to achieve further progress in this direction, it is imperative to establish the theoretical performance limits of this novel scheme and the associated design challenges - which is attempted in this manuscript. Specifically, here we show that - (i) for similar conditions, $DT$ assays can achieve two-three orders of magnitude lower detection limits, (ii) provide a novel detection methodology which requires no a priori information on reaction rates, (iii) extend the same methodology to back-extract the reaction rates, (iv) quantify the influence of system level parameters like the observation time and number of receptors on the limits of detection. Below, we first provide a direct comparison of $DT$ and $EP$ schemes which allows us to establish the above mentioned contributions.\\

\begin{figure}[thpb]
  \centering
    \includegraphics[width=0.5\textwidth]{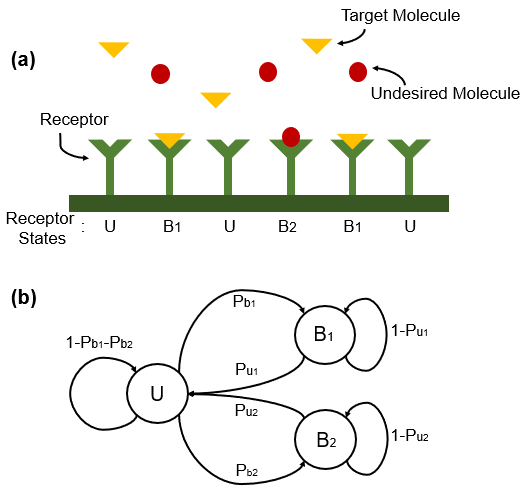}
    \caption{\textit{Schematic (a) and state transition diagram (b) to illustrate the receptor binding kinetics in biosensors. Here U represents an unbound receptor, $B_1$, a receptor bound to a target molecule and $B_2$, a receptor bound to an undesired molecule. The transition probabilities are dependent on the rate constants: $P_{b1} \propto k_{f1} \rho_{1}, \quad P_{b2} \propto k_{f2} \rho_{2}, \quad P_{u1} \propto k_{r1}, \quad P_{u2} \propto k_{r2}$. }}
\end{figure}

\section{\textbf{EP vs. DT sensors}}
 To obtain quantitative estimates on the performance of $DT$ and $EP$ sensors, we consider a sensor functionalized with receptors (density $N_0$) on its surface (see Fig. 1 a). The sensor is introduced to an analyte solution which contains both target (at concentration $\rho_1$) and undesired molecules (at a concentration $\rho_2$). The interactions of these molecules with the receptors are characterized by binding and unbinding affinities of $k_{f1}$ and $k_{r1}$ for the target and $k_{f2}$ and $k_{r2}$ for the undesired molecules.\vspace{-0.7em}\\
 
 The dynamics of the receptor conjugation by the target as well as the undesired molecules can be represented by state transitions. Each receptor molecule can exist in one of the three states: Unbound, Bound 1 (bound to a target molecule) and Bound 2 (bound to an undesired molecule). The different transition probabilities are governed by the concentrations as well as the affinities. (see Fig. 1 b)\vspace{-0.7em}\\
 
 From the state transition pattern, one can find $N_{1}$ and $N_{2}$, the number of receptor molecules bound to the target and undesired molecules, respectively, at equilibrium to be:
\begin{equation}
    \frac{N_{1}}{N_0} = \frac{k_{f1} \rho_1}{k_{f1} \rho_1 + k_{r1} + k_{r1} k_{f2} \rho_2/k_{r2}}
\end{equation}
\begin{equation}
    \frac{N_{2}}{N_0} = \frac{k_{f2} \rho_2}{k_{f2} \rho_2 + k_{r2} + k_{r2} k_{f1} \rho_1/k_{r1}}
\end{equation}
 \par The above equations and the state diagram help us evaluate the performance of $EP$ and $DT$ sensors. The $EP$ sensors rely on the steady state density of bound receptor molecules, $N_1 + N_2$, which can significantly depend on parameters related to the undesired molecules as evident from eqs. (1)-(2) above. On the other hand, the detection in a $DT$ scheme relies on the kinetics of receptor binding and unbinding events, which continue even at steady state. Thus, upon reaching steady state, the response of $EP$ sensors is expected to saturate (of course with some variation), while the $DT$ sensors continue to provide useful information in terms of dissociation kinetics which could result in sensitive detection.\vspace{-0.7em}\\
 
 For the end point assays, the signal is proportional to the total number of bound receptors $N_1 + N_2$, which is directly used to extract the target concentration, $\rho_1^{est}$. Accordingly, a simple analysis shows that
\begin{equation}
\frac{N_1 + N_2}{N_0} = \frac{k_{f1} \rho_1^{est}}{k_{f1} \rho_1^{est} + k_{r1}}.
\end{equation}
 \par The challenges posed by undesired molecules for $EP$ sensors is evident in eq. (3) as illustrated through the following example system: $\rho_1 = 10^{-9} M$, $\rho_2 = 10^{-7} M$, $k_{f1} = 10^7 M^{-1} s^{-1}$, $k_{r1} = 10^{-2}s^{-1}$, $k_{f2} = 10^7M^{-1} s^{-1}$, $k_{r2} = 1s^{-1}$, and $N_0 = 1000$.  In the absence of the undesired molecules, the steady state scenario would have 500 bound receptors. However, in the presence of undesired molecules we get $N_1 = 333$ and $N_2 = 333$  (using eqs (1)-(2), and the above listed parameters), resulting in a response which correspond to a total of 666 bound receptors. The back extracted $\rho_1^{est}$, obtained using eq. (3), would have an error of $100\%$, i.e. our prediction would be twice as much as the actual target density. In the presence of high abundance, low specificity molecules, endpoint assays give distorted results- both in terms of estimation of target molecule density and in terms of the absolute limits of detection\vspace{-0.7em}\\
 
 $DT$ sensors, on the other hand, rely on the kinetics of the dissociation of bound receptors to predict the target concentration. If we know the state of a receptor (whether bound or not - through some schemes like imaging as shown in \cite{Sevenler4129}), we need a strategy to distinguish between the target-receptor dissociation against target-undesired pair dissociation.  This is possible if there exists a significant difference in the unbinding time for the two. For example, under the given scenario of $k_{r1} = 0.01 s^{-1}$ and $k_{r2} = 1s^{-1}$, the average unbinding time would be $1/k_{r1} = 100 s$ for a target molecule and $1/k_{r2} = 1s$ for an undesired molecule. Once we obtain the information detailing the sensor events, the statistics of unbinding time for each bound molecule can be used to classify each of them as either a target molecule or an undesired molecule. For the above example, if we find that a molecule unbinds after 100s, it is more likely a target molecule than an undesired one. Similarly, if a molecule unbinds with a few seconds, it is more likely an undesired one than a target molecule. Of course, there might be undesired molecules that stick around for much longer than the average of 1s and there might be target molecules that unbind much faster than the average of 100s. We need to identify a threshold time to distinguish target vs. undesired dissociation events - a problem addressed in the next section.\\

\section{\textbf{Threshold time}}
It is evident that the choice of threshold time could indeed influence how the dissociation kinetics are analyzed and hence the estimation of target concentration and the limits of detection. The challenges in arriving at an estimate for the threshold time is two fold (i) a priori information on the dissociation time constants might not be available, (ii) the penalty or error associated with the choice of a given threshold time is not evident. Ideally, one would like to obtain an estimate for the threshold time under the scenario listed as item (i) while minimizing the error listed in item (ii). We first obtain thresholds for situations where $k_{r1}$ and $k_{r2}$ are known a priori. Later, we extend the arguments for the general case of item (i) as well.\vspace{-0.7em}\\

\textbf{Constrained Detection:} First, we develop an estimate for the threshold time under the assumption that all relevant reaction rates are known a priori. Under this scenario, a theoretical estimate for threshold time can be obtained by addressing the following question: Given that a molecule remained bound to the receptor for a total time of $t_{ub}$, what is the probability that the molecule was indeed the target? Such estimates are commonly addressed in probability theory and digital communication \cite{Gallager2008PrinciplesOD}, and a similar approach is useful here as well. For very short times, one can see that this probability would be almost 0, and for very long times, this probability would be almost 1. We define a time $t_{MAP}$ such that for $t_{ub} = t_{MAP}$, the unbinding event could have resulted equally from the states $B_1$ and $B_2$ (see the state diagram, Fig. 1). Or rather, in terms of conditional probabilities, $P(B_1 | t_{ub} = t_{MAP}) = P(B_2 | t_{ub} = t_{MAP}) = 0.5$. Under such a scenario, our optimal threshold is indeed $t_{MAP}$ and this is described as the Maximum a posteriori ($MAP$) estimation.\vspace{-0.7em}\\

Another way to look at this problem is to ask the question: Which among the target and undesired molecules is more likely to have an unbinding time of $t_{ub}$? And to find the corresponding threshold $t_{ML}$, we would have the condition $P(t_{ub} = t_{ML} | B_1) = P(t_{ub} = t_{ML} | B_2)$. This technique is referred to as the Maximum Likelihood ($ML$) estimation.\vspace{-0.7em}\\

The unbinding time has an exponential probability distribution of $P(t_{ub}|B_1) = k_{r1} e^{-k_{r1} t_{ub}}$. The ML threshold, $t_{ML}$, is obtained by setting $k_{r1} e^{-k_{r1} t_{ML}} = k_{r2} e^{-k_{r2} t_{ML}}$, which leads to
\begin{equation}
    t_{ML} = \frac{ln(k_{r2}) - ln(k_{r1})}{k_{r2}-k_{r1}}
\end{equation}
 
\par On the other hand, MAP estimation is a bit more involved. As mentioned before, the criteria for $MAP$ estimation is $P(B_1 | t_{ub} = t_{MAP}) = P(B_2 | t_{ub} = t_{MAP})$. This expression involves conditional probabilities and can be reduced to $P(t_{ub} = t_{MAP}|B_1) P(B_1) = P(t_{ub} = t_{MAP}|B_2) P(B_2)$. From the state transitions (see Fig. 1 b), we note that the prior probability, $P(\text{unbind from }B_{1}) \propto N_1 k_{r1}$. The $MAP$ criteria can then be reduced to $N_1 k_{r1}^2 e^{-k_{r1} t_{MAP}}  = N_2 k_{r2}^2 e^{-k_{r2} t_{MAP}}$ which leads to   
\begin{equation}
    t_{MAP} = \frac{2\big(ln(k_{r2}) - ln(k_{r1})\big) + ln(N_{2}) - ln(N_{1})}{k_{r2}-k_{r1}}
\end{equation}

\par Note that MAP estimate requires a priori knowledge of the parameters $N_1$ and $N_2$ and the dissociation rates, while the ML estimate requires a priori information of only the dissociation rates. Hence, the MAP threshold cannot be computed directly. For the specific example considered in section \uppercase\expandafter{\romannumeral2\relax}, the computed thresholds values are $t_{ML} = 4.65s$ and $t_{MAP} = 9.3s$\vspace{-0.7em}\\

To justify the discussion of thresholds presented above, we performed detailed statistical simulations. First, we simulated the receptor binding-unbinding events, for a given set of concentrations and affinities over an observation time of T = 2000s, as per the state transition diagram considered (Fig. 1 a). For each receptor, we kept track of the binding/unbinding events, without making the distinction of whether the binding/unbinding species is a target or an undesired molecule. Next, we used a threshold, $t_{thresh}$, to classify each unbinding species as either a target or an undesired molecule. The molecules that unbind after a time longer than $t_{thresh}$ are considered as target molecules, while those that unbind sooner are treated as undesired molecules. Performing this analysis on each unbinding event, leaves us with a reconstruction of the states of each receptor at each time instant. We thus have predictions on the total number of receptors bound to target molecules ($N_1$) and the total number of receptors bound to undesired molecules ($N_2$) in steady state. Eq. (1)-(2) were then used to predict the target concentration, $\rho_1^{est}$ and undesired molecule concentration, $\rho_2^{est}$ for one choice of $t_{thresh}$. The threshold time, $t_{thresh}$, was varied, and for each threshold, we get different event reconstructions, and hence different estimations $\rho_1^{est}$ and $\rho_2^{est}$.\vspace{-0.7em}\\ 

The optimal threshold must then be the threshold value where the estimated concentrations match the actual concentration. The results shown in Fig. 2 clearly indicate that the optimal threshold is indeed the MAP threshold described above. For each reconstruction, we also obtain average unbinding times for target molecules and undesired molecules. The estimates of dissociation rates ($k_{r1}^{est}$ and $k_{r2}^{est}$) are then simply computed as the inverse of the average unbinding times. These estimates, for different threshold values, are shown in Fig. 3. These too match the actual values when the MAP threshold is chosen. If the dissociation rates are known a priori, one can set up the optimal MAP threshold by simply varying it, and checking for a match between the estimated and actual dissociation rate. Once the MAP threshold is obtained by matching the dissociation rates, it can then be used to find the concentration.\vspace{-0.7em}\\
\begin{figure}[h!]
  \centering
    \includegraphics[width=0.45\textwidth]{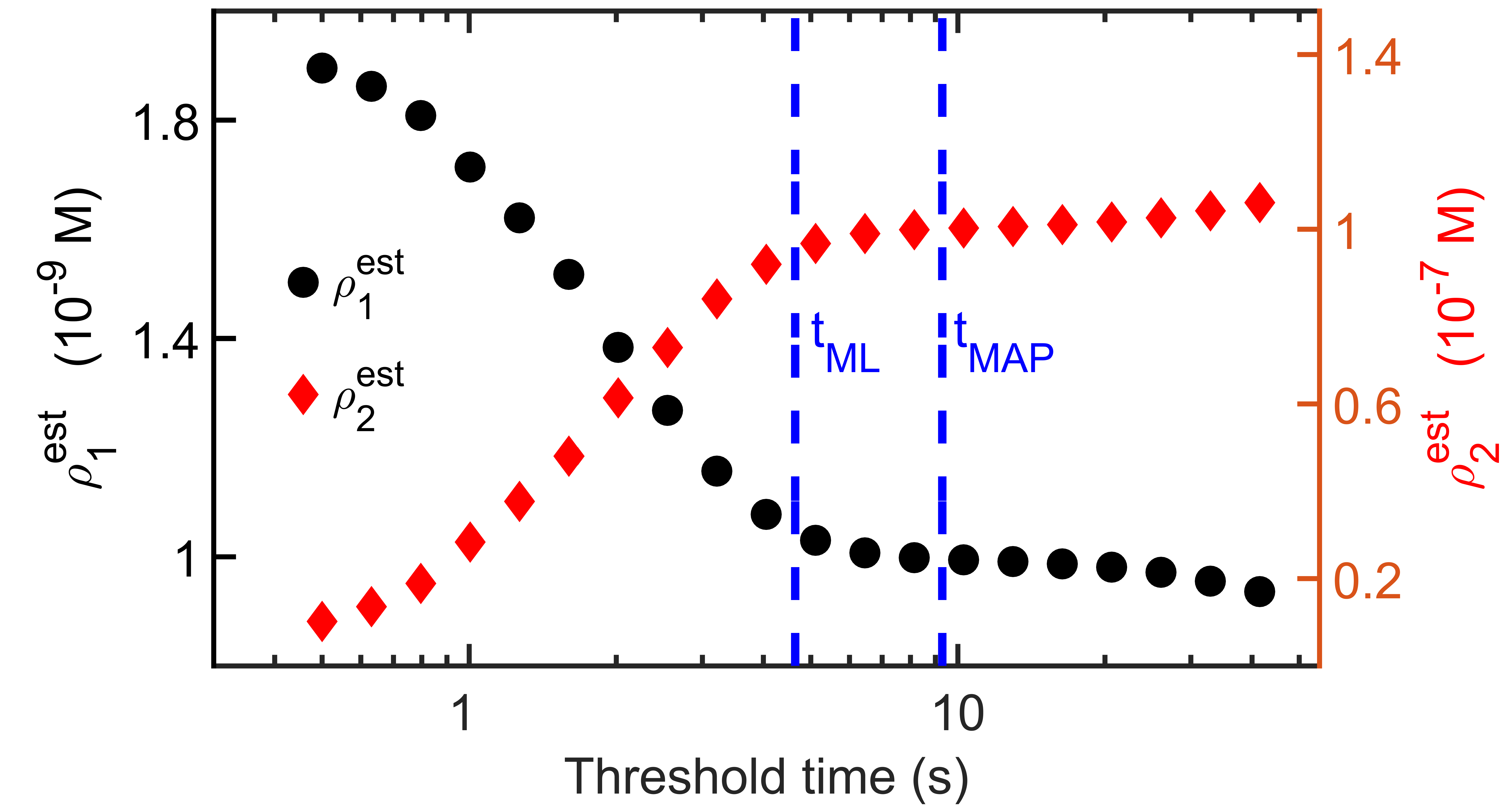}
    \caption{\textit{Estimated concentrations vs. threshold time}}
\end{figure}   
\begin{figure}[h!]
  \centering
    \includegraphics[width=0.45\textwidth]{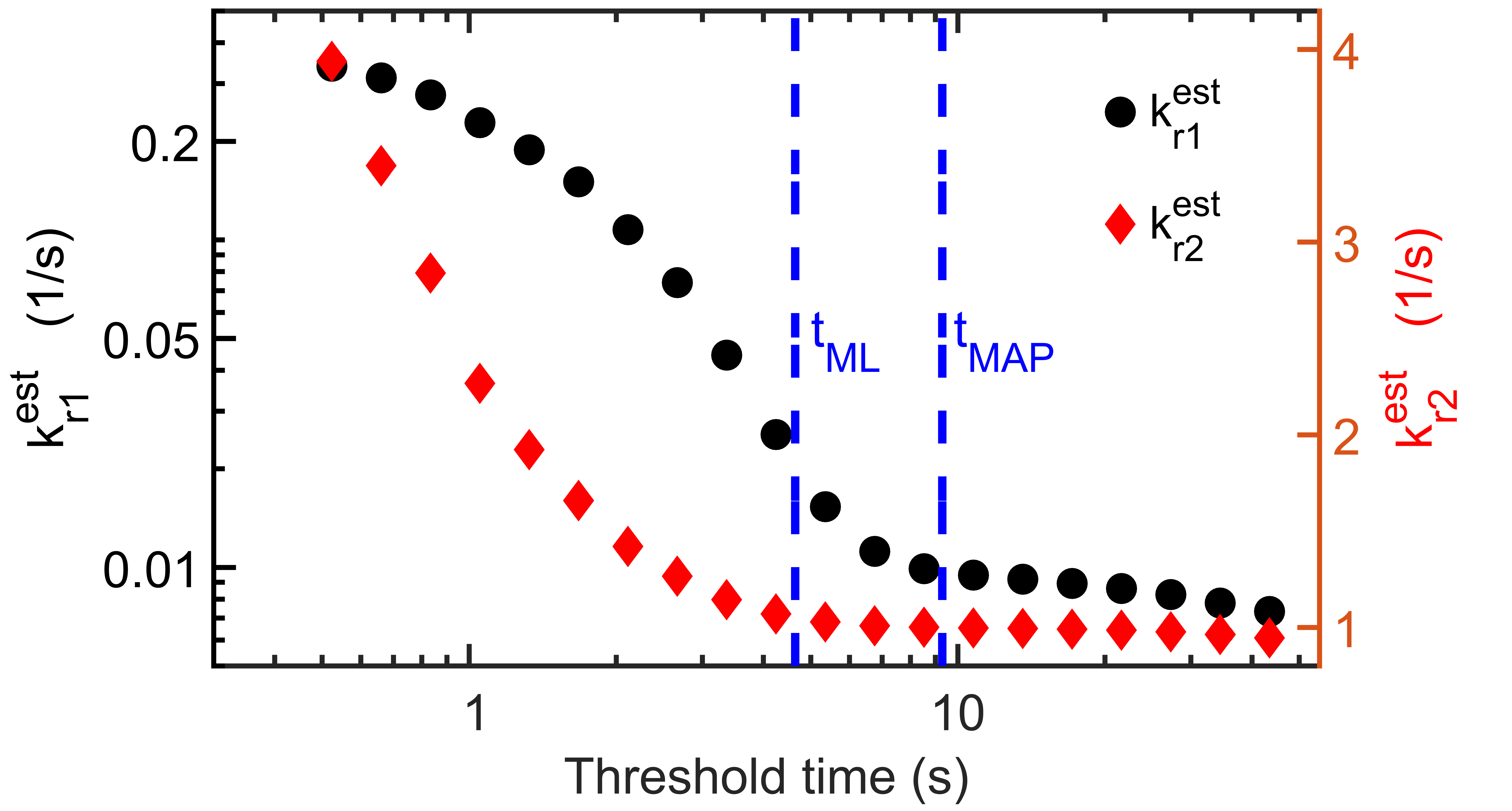}
    \caption{\textit{Estimated dissociation rates vs. threshold time}}
\end{figure}   

\par \textbf{Unconstrained detection:} The above analysis indicates that the threshold time used to identify target vs. undesired dissociation events is indeed the $t_{MAP}$. We further note that although not optimal, the $t_{ML}$ would provide good estimates as well. However, both these require a priori information on the dissociation constants and hence are not equipped to address the challenge listed as item (i) at the beginning of this section i.e., to devise a detection strategy when prior information is not available (Unconstrained detection). Interestingly, the results shown in Fig. 2 and 3 resolve this in a simple manner. We see that the target dissociation rate estimate ($k_{r1}^{est}$) in Fig. 3 has a `knee' (a sudden change in slope) at the MAP threshold. This observation enables the following detection strategy: After collecting data on the unbinding events, we proceed as before, and estimate the concentrations and dissociation rates for different choices of thresholds. We then identify the MAP threshold as the knee-region in the $k_{r1}^{est}$ curve, and this threshold is used to obtain the best estimate of the target concentration.\vspace{-0.7em}\\


Our proposed unconstrained detection scheme is significant in a variety of aspects as listed: (i) Often the reaction constants are obtained through a solution based scheme and the effective reaction constants for receptors bound on a surface could be significantly different. In such cases, the unconstrained detection scheme is immensely useful. (ii) the proposed scheme, in conjunction with eqs. (4)-(5) allows back extraction of critical parameters like the reaction rates and the $N_0$ (see Fig. 3). (iii) Even if the dissociation constants $k_{r1}$ and $k_{r2}$ are not too distinct, one can employ correction schemes or measurement protocols to improve the estimates (see supplementary materials).

\section{\textbf{Uncertainty in detection}}
\par In the previous section, we proposed and numerically validated a scheme to achieve detection even when the dissociation rates are not a priori known. An associated important figure of merit is the error or uncertainty involved in the detection. It is evident that the uncertainty is influenced by the time duration ($T$) over which data is collected. The duration should be long enough to ensure that enough unbinding events are observed which could provide an accurate estimate for the target density. Similarly, the number of receptors ($N_0$) could also influence the detection scheme. We first quantify an estimate for the uncertainty involved in this scheme and later use the same to obtain the performance limits.\vspace{0.3 em}

As seen in the previous section, the performance of the proposed scheme depends significantly on the threshold time we use. In order to set up an optimal threshold time, we estimated the dissociation rates for different threshold values and chose the threshold at which our estimate matched closely with the expected value. The estimate of dissociation rate was obtained as the inverse of the average unbinding time. The finite observation duration of $T$, means that we observe only a finite number of unbinding events with which we compute this average. We hence expect some level of uncertainty in this estimate. The inverse relationship between the average unbinding time and the dissociation rate means that a small uncertainty, say of 5\%, in the average unbinding time, would translate to an almost same, 5\%, uncertainty in the dissociation rate estimate.\vspace{-0.7em}\\

To estimate the uncertainty involved, we consider the set of all the target unbinding events that happen within the total observation time ($T$). Let $N_{ub}$ denote the number of these events over the entire duration and $\mu$ and $\sigma$ denote the mean and standard deviation of the unbinding times, respectively. Given this, central limit theorem \cite{william2014statistics} indicates that the uncertainty in the estimated dissociation time of $DT$ sensors, $U_{DT}$ is 

\begin{equation}
U_{DT}=\frac{\sigma}{\mu\sqrt{N_{ub}}}
\end{equation}

\par We know that the dissociation time follows an exponential probability distribution and for the target unbinding time, we have: $p(t) = k_{r1} e^{-k_{r1}t}$. For an observation time of $T$, considering a small threshold, the mean ($\mu$) and variance ($\sigma^2$) of expected unbinding time is given by
\begin{equation}
\mu = \sigma = 1/k_{r1}.
\end{equation}

\par From the state transition considered, we know that the probability that a bound target molecule unbinds is proportional to $k_{r1}$. For a total of $N_1$ such receptors bound in equilibrium, and over time $T$, we would have a total of $N_{ub} = N_1 k_{r1} T$ unbinding events. We also have $N_1 = \alpha N_0$, where $\alpha$ is a function of concentrations and affinities, given in eq. (1). Hence,
\begin{equation}
    N_{ub} = \alpha N_0 k_{r1} T.
\end{equation}

\par Using eqs. (6)-(8), we find that the uncertainty associated with the $DT$ scheme as

\begin{equation}
U_{DT} = \frac{1}{\sqrt{\alpha N_0 k_{r1} T}}.
\end{equation}

\par A striking feature of eq (9) is the inverse square relation of the error with both the total observation time as well as the number of receptors. This is expected because in a dynamic tracking biosensor, unlike a traditional assay, we keep getting useful information even after reaching equilibrium. We are able to improve the accuracy of the target concentration prediction by increasing the observed number of target-receptor events - through both the observation time window and number of receptors.\\

\section{\textbf{Performance Limits}} 
Having obtained an estimate for the uncertainty associated with $DT$ sensors, we now attempt to evaluate its performance limits and compare the same with $EP$ sensors. Using eq. (1)-(3), one can show that the error in the target concentration prediction for a traditional $EP$ biosensor in the presence of undesired molecules is given by:
\begin{equation}
U_{EP}= \frac{\rho_1^{est} - \rho_1}{\rho_1} = \frac{k_{f2} \rho_{2} / k_{r2}}{k_{f1} \rho_{1} /k_{r1} } = \frac{x_2}{x_1}
\end{equation}
\par Here we have defined quantities $x_1 = k_{f1} \rho_{1} / k_{r1}$ and $x_2 = k_{f2} \rho_{2} / k_{r2}$. One can observe that for low concentrations or high dissociation rates of undesired molecules, the error is quite small, but as the concentration of undesired molecules rises, the traditional biosensor becomes unusable.\vspace{-0.7em}\\

For $DT$ sensors, using eq. (9), the associated error in terms of $x_1$ and $x_2$, is
\begin{equation}
U_{DT} = \frac{1}{\sqrt{N_0 k_{r1} T}} \sqrt{\frac{1+x_1 +x_2}{x_1}}
\end{equation}

The errors of both, the dynamic tracking biosensor and the traditional biosensor, have a dependence on the quantity $x_2 = k_{f2} \rho_{2} / k_{r2}$. In essence, this quantity can be viewed as a `normalized' concentration of undesired molecules. It is also clear that the larger this quantity is, the larger is the error in both cases. For a given value of $x_2$, to obtain a reliable target concentration prediction, we would need to ensure that the prediction error is low enough, say 5\%. This sets a lower bound on the target concentration that can be detected.\vspace{-0.7em}\\

 Fig. 4 presents a comparison of the minimum detectable target concentration and the region of operation of both schemes. We have considered the target-receptor dissociation constant to be $k_{r1} = 0.01 s^{-1}$, the number of receptors to be $N_0 = 1000$ and the total observation time to be $T=2000 s$. We have both the target concentration as well as the undesired molecule concentration in the `normalized' units, $x_1$ and $x_2$. In such a case, using a $DT$ biosensor, we see an improvement of two orders in magnitude in the minimum detectable target concentration (at most 5\% error) - as compared to $EP$.\vspace{-0.7em}\\
 
 Fig. 5 shows the performance of a dynamic tracking biosensor for three cases: i) $N=100$, $T=2000$, ii) $N=1000$, $T=2000$ and iii) $N=1000$, $T=20000$. We see an improvement in performance with both, an increase in $N$ and an increase in $T$, as predicted by eq. (9)\vspace{-0.7em}\\
 
\section{\textbf{Conclusions}}
In summary, here we provided a detailed theoretical analysis on dynamic tracking biosensors. Based on the state transition diagram, we first established the appropriate detection scheme when all parameters are known. We then proposed a methodology to achieve sensitive detection even when the reaction parameters are not known a priori. This analysis enabled us to compare and contrast the $EP$ and $DT$ sensors in terms of their performance limits. Curiously, well designed $DT$ schemes can be $2-3$ orders of magnitude more sensitive than $EP$ sensors. Finally, we elucidated the functional dependence of $DT$ sensors on various critical parameters like the receptor density and observation time window. As such, this detailed analysis should enable careful design of experiments to achieve the promising potential of $DT$ sensors towards ultra sensitive detection of biomolecules.

\newpage 

\begin{figure}[h!]
  \centering
    \includegraphics[width=0.45\textwidth]{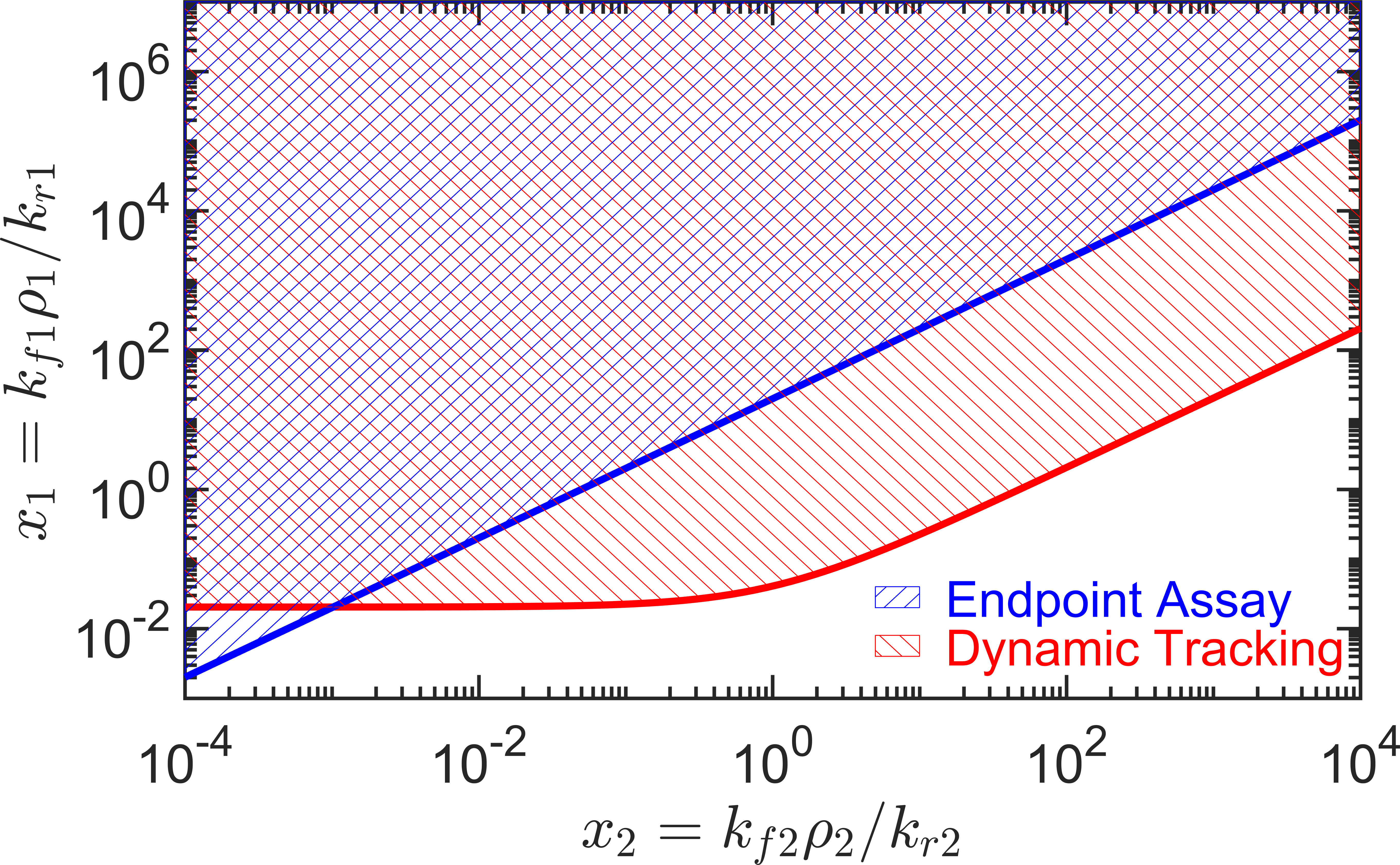}
    \caption{\textit{Region of operation of DT vs EP Biosensor}}
\end{figure}   

\begin{figure}[h!]
  \centering
    \includegraphics[width=0.45\textwidth]{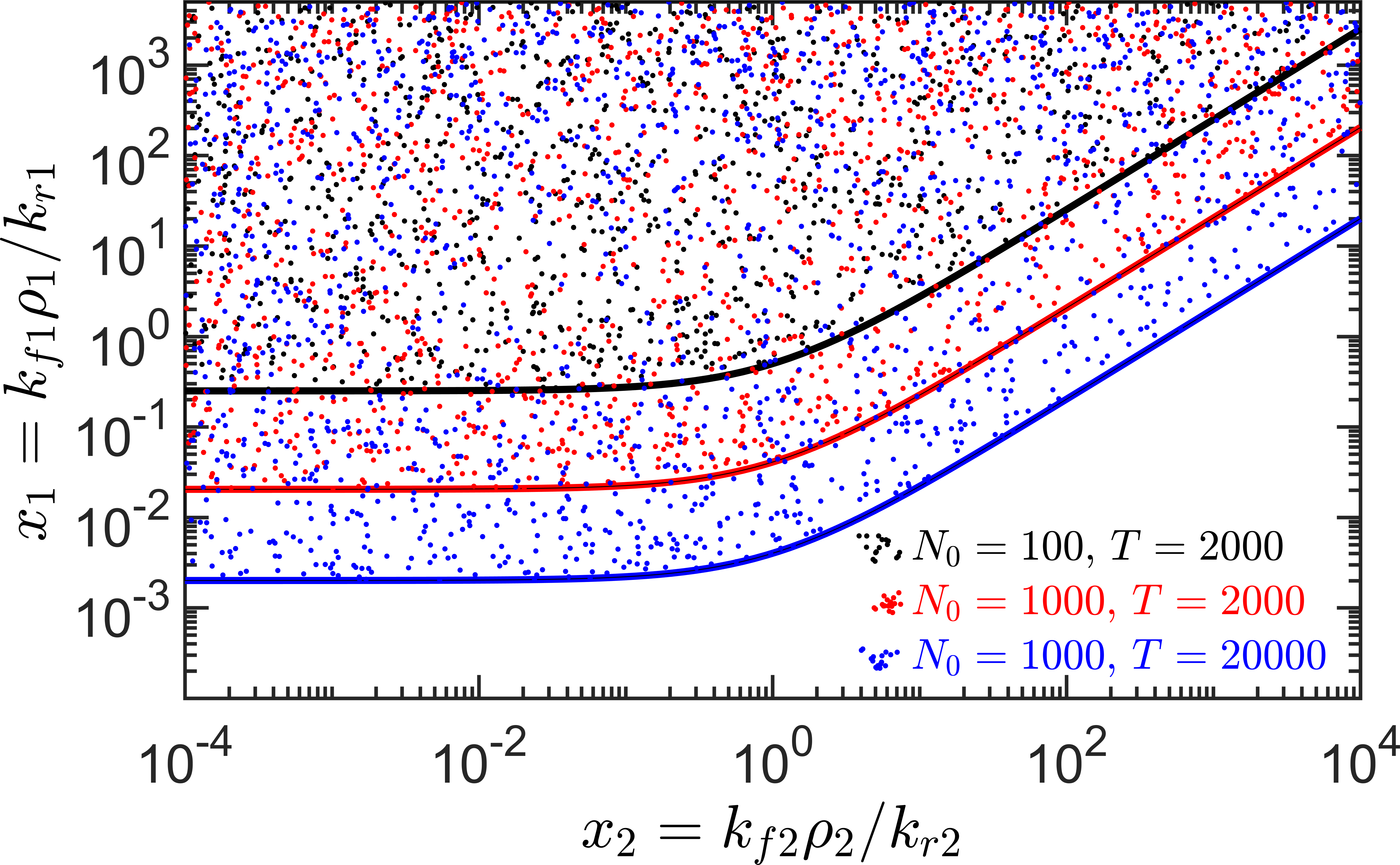}
    \captionsetup{justification=centering}
    \caption{\textit{Region of operation of DT Biosensor \newline for different $N_0$ and $T$}}
\end{figure} 

\addtolength{\textheight}{-12cm}   



\bibliography{ref2.bib}{}\bibliographystyle{IEEEtran}

\end{document}